\documentclass{aa}
\usepackage{graphicx}
\usepackage{txfonts}

\begin{document}

\title{Hints for a fast precessing relativistic radio jet in
LS~I~+61$^{\circ}$303}

\author{M. Massi\inst{1}
\and M. Rib\'o\inst{2}
\and J.~M. Paredes\inst{3,}\thanks{CER on Astrophysics, Particle Physics
and Cosmology. Universitat de Barcelona}
\and S.~T.\ Garrington\inst{4}
\and M. Peracaula\inst{5}
\and J. Mart\'{\i}\inst{6}
}

\institute{Max Planck Institut f\"ur Radioastronomie, Auf dem H\"ugel 69, 53121 Bonn, Germany\\
\email{mmassi@mpifr-bonn.mpg.de}
\and Service d'Astrophysique, CEA Saclay, B\^at. 709, L'Orme des Merisiers, 91191 Gif-sur-Yvette, Cedex, France\\
\email{mribo@discovery.saclay.cea.fr}
\and Departament d'Astronomia i Meteorologia, Universitat de Barcelona, Av. Diagonal 647, 08028 Barcelona, Spain\\
\email{jmparedes@ub.edu}
\and Nuffield Radio Astronomy Laboratories, Jodrell Bank, Macclesfield, Cheshire SK11 9DL, UK\\
\email{stg@jb.man.ac.uk}
\and Institut d'Inform\`atica i Aplicacions, Universitat de Girona, Campus de
Montilivi s/n, 17071 Girona, Spain\\
\email{marta.peracaula@udg.es}
\and Departamento de F\'{\i}sica, Escuela Polit\'ecnica Superior, Univ. de Ja\'en, Virgen de la Cabeza 2, 23071 Ja\'en, Spain\\
\email{jmarti@ujaen.es}
} 

\offprints{M. Massi, \\ \email{mmassi@mpifr-bonn.mpg.de}}

\date{Received / Accepted}

\abstract{
Here we discuss two consecutive MERLIN observations of the X-ray binary 
\object{LS~I~+61$^{\circ}$303}. The first observation shows a double-sided jet
extending up to about 200~AU on both sides of a central source. The jet shows
a bent S-shaped structure similar to the one displayed by the well-known
precessing jet of \object{SS~433}. The precession suggested in the first
MERLIN image becomes evident in the second one, showing a one-sided bent jet
significantly rotated with respect to the jet of the day before. We conclude
that the derived precession of the relativistic ($\beta$=0.6) jet explains
puzzling previous VLBI results. Moreover, the fact that the precession is fast
could be the explanation of the never understood short term (days) variability
of the associated gamma-ray source
\object{2CG~135+01}/\object{3EG~J0241+6103}.
}

\maketitle

\section{Introduction} \label{introduction}

In 1978 Gregory and Taylor reported the discovery of a highly variable radio
source within the 1$\sigma$ error circle of the $\gamma$-ray source 2CG~135+01
(Gregory \& Taylor \cite{gregory78}). The radio source turned out to be
periodic, with a periodicity of about 26 days and coincident with the stellar
binary system \object{LS~I~+61$^{\circ}$303} (Taylor et~al. \cite{taylor80};
Hjellming et~al. \cite{hjellming78}; Gregory \cite{gregory02}).

In 1993 Massi and collaborators showed by a Very Long Baseline Interferometry
(VLBI) observation that the radio emission had a structure of milliarcsecond
(mas) size corresponding to a few AU at the distance of 2.0~kpc (Massi et~al.
\cite{massi93}, Frail \& Hjellming \cite{frail91}). Whereas the evidence of
such a structure included \object{LS~I~+61$^{\circ}$303} in the small subclass
of X-ray binaries having associated relativistic radio-jets (Fender et~al.
\cite{fender97}), systems now generally called microquasars (Mirabel \&
Rodr\'{\i}guez \cite{mirabel99}; Fender \cite{fender03}), the complex
morphology in this and successive VLBI observations made an interpretation in
terms of a collimated ejection with a constant position angle difficult
(Peracaula et~al. \cite{peracaula98}; Paredes et~al. \cite{paredes98}; Taylor
et~al. \cite{taylor00}). The turning point came from an image at a larger
scale (up to tens of AU) performed with part of the European VLBI Network
(EVN). This showed an elongation in a clear direction, interpreted as a
one-sided Doppler-boosted jet (Massi et~al. \cite{massi01}). Thus the very
confusing structure close to the binary system seemed to be better
distinguished at a larger distance from the core. Consequently, we decided to
explore \object{LS~I~+61$^{\circ}$303} at even larger scales (up to hundreds
of AU) using the Multi-Element Radio-Linked Interferometer Network (MERLIN).

\section{Observations and results} \label{observations}

\begin{figure*}[t!]
\begin{center}
\resizebox{1.0\hsize}{!}{\includegraphics{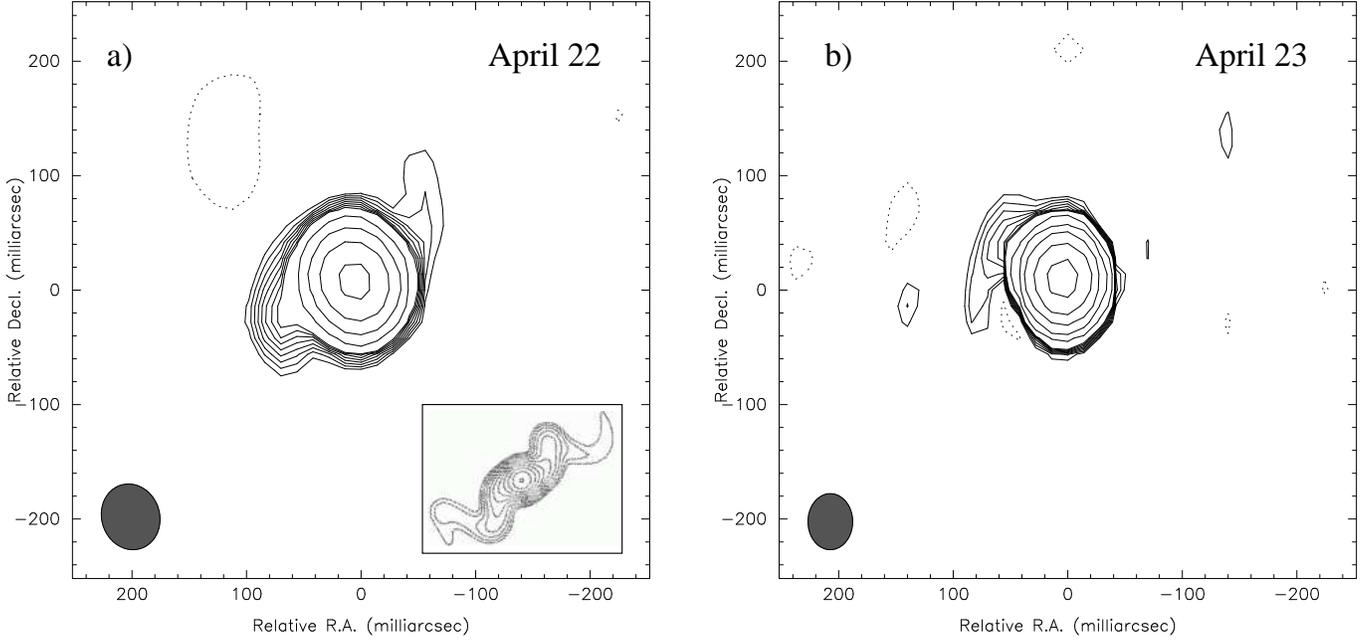}}
\caption{
{\bf a)} MERLIN self-calibrated image of \object{LS~I~+61$^{\circ}$303} at
5~GHz and using natural weights, obtained on 2001 April 22. North is up and
East is to the left. The synthesized beam has a size of $51\times58$~mas, with
a PA of 17\degr. The contour levels are at $-$3, 3, 4, 5, 6, 7, 8, 9, 10, 20,
40, 80, and 160$\sigma$, being $\sigma$=0.14~mJy~beam$^{-1}$. The S-shaped
morphology strongly recalls the precessing jet of \object{SS~433}, whose
simulated radio emission (Fig.~6b in Hjellming \& Johnston \cite{hjellming88},
rotated here for comparison purposes) is given in the small box. {\bf b)} Same
as before but for the April 23 run and using uniform weights (see text). The
synthesized beam has a size of $39\times49$~mas, with a PA of $-$10\degr. The
contour levels are the same as those used in the April 22 image but up to
320$\sigma$, with $\sigma$=0.12~mJy~beam$^{-1}$.
}
\label{fig:merlin}
\end{center}
\end{figure*}

We performed the observations of \object{LS~I~+61$^{\circ}$303} with MERLIN at
5~GHz on 2001 April 22 and April 23. The log of the observation is given in
Table~\ref{table:log}. The total bandwidth was 32~MHz, the sampling was 2-bit
and the correlator integration time 8~s. The data were calibrated using the
pipeline available at Jodrell Bank Observatory within the {\sc aips} software
package. The image processing was performed with {\sc difmap}.

\begin{table}
\caption[]{Log of the MERLIN observations. Start and Stop are given in Modified Julian Date (MJD=JD$-$2400000.5). The corresponding orbital phases have been computed using the new ephemerides, $t_0$=JD\,2443366.775 and $P$=26.4960~d, from Gregory (\cite{gregory02}).}
\begin{center}
\begin{tabular}{lcccc}
\hline \hline \noalign{\smallskip}
Date     & Start MJD & Stop MJD & $\phi_{\rm start}$ & $\phi_{\rm stop}$ \\
\noalign{\smallskip} \hline \noalign{\smallskip}
April 22 & 52021.73  & 52022.10 & 0.670              & 0.684 \\
April 23 & 52022.68  & 52023.17 & 0.706              & 0.724 \\
\noalign{\smallskip} \hline
\end{tabular}
\end{center}
\label{table:log}
\end{table}

\begin{table}
\caption[]{Parameters of the jet components. The distance and PA are relative to the Core, which is not at the phase center in Fig~\ref{fig:merlin}.}
\begin{center}
\begin{tabular}{llccc}
\hline \hline \noalign{\smallskip}
Date     & Component   &  Flux  (mJy) & Distance (mas) & PA (\degr)\\
\noalign{\smallskip} \hline \noalign{\smallskip}
April 22 & Core        &  28.0        & ---            & --- \\
         & South-East  &  ~~1.8       & 73             &   116 \\
         & North-West  &  ~~1.0       & 75             & $-$48 \\
\noalign{\smallskip} \hline \noalign{\smallskip}
April 23 & Core        &  51.4        & ---            & --- \\
         & North-East  &  ~~1.1       & 61             &   ~67 \\
\noalign{\smallskip} \hline
\end{tabular}
\end{center}
\label{table:components}
\end{table}

The image for April 22, shown in Fig.~\ref{fig:merlin}a, has been made with
natural weights in order to enhance faint and extended structures. The
morphology of the radio emission is a two-sided jet emanating from a central
core. The fluxes of the components are given in Table~\ref{table:components}.
The jet has a total length of $\sim$200~mas ($\sim$400~AU at a distance of
2.0~kpc) and the line joining the two lobes has a position angle (PA) of
$124\pm16\degr$. The morphology of the MERLIN image has a bent, S-like
structure. We show in the small box in Fig.~\ref{fig:merlin}a the simulated
radio emission by the Hjellming \& Johnston (\cite{hjellming88}) model of the
precessing jet of \object{SS~433} (rotated here for comparison purposes). The
similarity between the MERLIN image of \object{LS~I~+61$^{\circ}$303} and the
precessing model for \object{SS~433} suggests a precession of the jet.

\begin{figure}[t!]
\begin{center}
\resizebox{1.0\hsize}{!}{\includegraphics{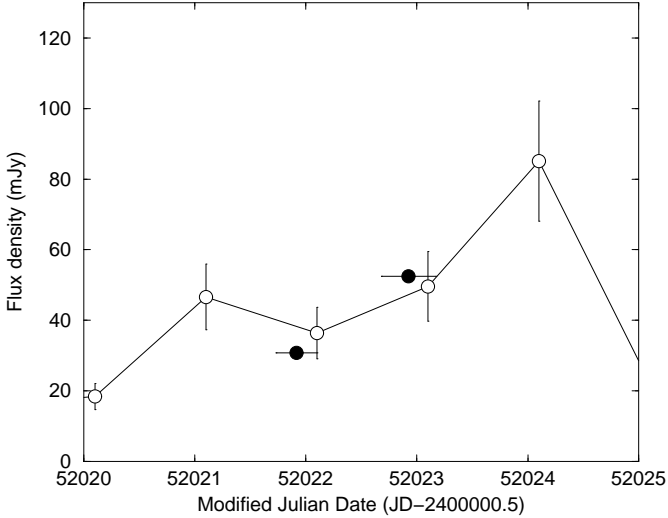}}
\caption{Lightcurve of \object{LS~I~+61$^{\circ}$303} at 3.9~GHz obtained with RATAN-600 (open circles). Indicative error bars of 20\% of the flux density have been plotted. The filled circles represent the MERLIN flux densities of the structures quoted in Table~\ref{table:components}, while the horizontal bars extend from the start to the stop of each observing run.}
\label{fig:ratan}
\end{center}
\end{figure}

\begin{figure*}[t!]
\begin{center}
\resizebox{1.0\hsize}{!}{\includegraphics{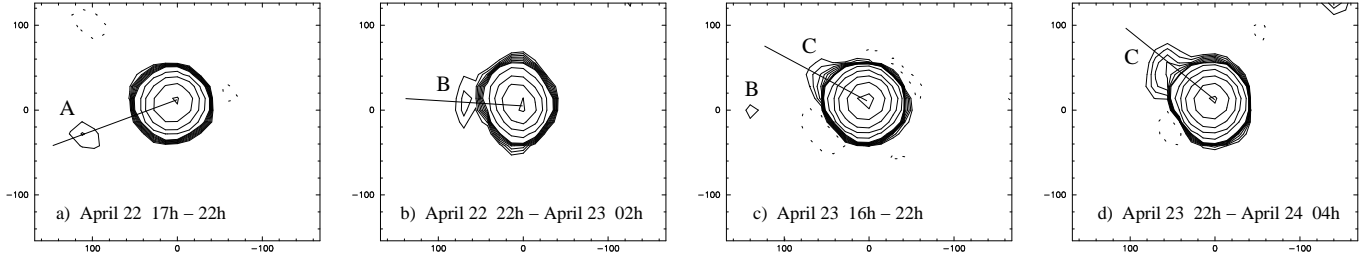}}
\caption{
MERLIN self-calibrated images of \object{LS~I~+61$^{\circ}$303} at 5~GHz using uniform weights, obtained on 2001 April 22 and April 23. The data set of each epoch has been split into two blocks. A convolving beam of 40~mas has been used in all images for better display. The first contour represents the $3\sigma$ level in all images except for c), where we start from the $2\sigma$ level to display the faint B component. The rms noises are $\sigma$=0.13~mJy~beam$^{-1}$, $\sigma$=0.20~mJy~beam$^{-1}$, $\sigma$=0.13~mJy~beam$^{-1}$, and $\sigma$=0.15~mJy~beam$^{-1}$, respectively. The PA of the ejections is indicated by a bar (see text).
}
\label{fig:four}
\end{center}
\end{figure*}

The precession suggested in the first MERLIN image becomes evident in the
second one, shown in Fig.~\ref{fig:merlin}b, where a new feature is present
oriented to North-East at a position angle (PA) of 67$\degr$. The 
Northwest-Southeast jet of Fig.~\ref{fig:merlin}a has a PA=124$\degr$.
Therefore a quite large rotation has occurred in only 24 hours. This fast
precession causes a deformation of the morphology during the second
observation, and the one-sided jet appears bent in Fig~\ref{fig:merlin}b. Only
3$\sigma$ features can be associated with the double jet of the day before,
the feature at 3$\sigma$ to the East is well compatible with a displacement of
$0.6c\times24$ hours (see discussion on the jet velocity in
Sect.~\ref{doppler}).

The MERLIN images could be affected to some degree by the variations in the
source structure and brightness during the 9 and 12 hour time span of each
observation, respectively. We show in Fig.~\ref{fig:ratan} the RATAN-600 data
at 3.9~GHz (monitoring program of microquasars by Trushkin et~al.
\cite{trushkin01}), together with our MERLIN flux density measurement at
5~GHz. As can be seen, the April 22 observation was performed during the
decaying phase of a small outburst, whereas the second run was performed
during the rising phase of a new ejection. While these variations would affect
high resolution VLBI observations, at the low MERLIN resolutions (as pointed
out by Fender et~al. \cite{fender99}) flux and structural variations would
only increase uncertainties in the position of the components (at the level of
a few mas) and in their flux density (of a few percent of their peak).

We have split the data of each epoch into two blocks and imaged them
separately (Fig.~\ref{fig:four}). We see that the Eastern bent structure
present in Fig.~\ref{fig:merlin}a is the result of a combination of an old
ejection A (Fig.~\ref{fig:four}a), already displaced 120~mas from the core,
and a new ejection B (Fig.~\ref{fig:four}b). After 19 hours
(Fig.~\ref{fig:four}c) the feature B is reduced at $2\sigma$ and the new
ejection C, at a clearly different PA with respect to B, is present. In
Fig.~\ref{fig:four}d, 6 hours later, little rotation of the PA is compatible
with $\Delta$ PA$_{(\rm B-\rm C)}/ 3$ of the previous image.

\section{Variable Doppler boosting} \label{doppler}


Position angle PA and $\theta$, the angle between the jet and the line of
sight, may both change as a function of time due to precession. It is $\theta$
the angle of physical relevance, because it influences the observed flux
density of the approaching ($S_{\rm a}$) and the receding ($S_{\rm r}$) jet
through the Doppler factor: $\delta_{\rm a,r}=[ \Gamma (1 \mp
\beta\cos\theta)]^{-1}$, where $\Gamma=(1-\beta^2)^{-1/2}$ is the Lorentz
factor and $\beta~c$ the jet velocity. The observed flux density of the
approaching and receding jets will be boosted and de-boosted respectively as
$S_{\rm a,r}=S \delta_{\rm a,r}^{k-\alpha}$, where $\alpha$ is the spectral
index of the emission ($S_{\nu}\propto \nu^{+\alpha}$) and $k$ is 2 for a
continuous jet and 3 for discrete condensations. The resulting ratio, given by
${S_{\rm a}\over{S_{\rm r}}}=\left({1+\beta\cos\theta\over1-\beta\cos\theta}
\right)^{k-\alpha}$, implies that for relativistic jets only for large
$\theta$ values will the two fluxes be comparable (as seems to be the case in
Fig.~\ref{fig:merlin}a).


As proved for a case of multiple ejections in \object{GRS~1915+105} (Fender
et~al. \cite{fender99}) the value of $k$ for each single ejection should be 
close to 2. This implies that the discrete components might be just the bright
parts of a continuous jet (Fender et~al. \cite{fender99}). Regarding $\alpha$,
\object{LS~I~+61$^{\circ}$303} is always optically thin at frequencies
(5--9)~GHz, even during the onset of radio outbursts (Strickman et~al.
\cite{strickman98}), and remains quite constant at a value of $-$0.5.

In Massi et~al. (\cite{massi01}) we used the values $\alpha=-0.5$ and $k=3$
resulting in $\beta>0.4$ and $\theta\sim0\degr$ for the EVN data. By adopting 
the more convincing value of $k=2$ the value of $\beta$ derived from the EVN
image becomes 0.6, while $\theta$ remains approximately zero. In the case of
the MERLIN image of April 22 we derive $\beta\cos\theta=0.12$, which for
$\beta=0.6$ leads to an ejection angle of $\theta=78\degr$. This is an average
of the ejection angles $\theta_{\rm A }$ and $\theta_{\rm B}$ of features A
and B in Figs.~\ref{fig:four}a and \ref{fig:four}b. A direct estimate of these
angles is prevented by the lack of the receding jets. By using the r.m.s.
noise we derive $\theta_{\rm A} < 90\degr$, $\theta_{\rm B}< 80\degr$ and for
the C ejection in Fig.~\ref{fig:four}c, $\theta_{\rm C} < 68 \degr$.

\section{Conclusions and discussion} \label{discussion}

We have shown, for the first time, the presence of a double-sided jet in 
\object{LS~I~+61$^{\circ}$303}.

The comparison of the MERLIN 2001 April 22 observation with our previous EVN
results allows us to establish a variation of the angle ($\theta$) between the
jet and the line of sight up to 78\degr between the two epochs. The variation
is attributable to a precession of the jet. Because of such a precession the
position angle (PA) of the projection of the jet onto the sky changes as well.
This explains the different alignments of PA$\simeq 30\degr$ or PA$\simeq
120$--$160\degr$ measured in different epochs (see Table~2 in Massi et~al.
\cite{massi01}).

The MERLIN 2001 April 23 image confirms a precession of the jet by the large
variation of both the values of PA and of $\theta$. In addition it reveals
quite short time-scales for the precession, since the time interval between
the two observations is of only 24 hours.

\object{LS~I~+61$^{\circ}$303} coincides with the high-energy gamma-ray source
\object{2CG~135+01} (\object{3EG~J0241+6103}) (Hartman et~al.
\cite{hartman99}). The emission could consist of inverse Compton upscattered
UV photons from the stellar companion by the relativistic electrons of the jet
as discussed by Taylor et~al. (\cite{taylor96}). In this case the fast
precession, pointing the jet intermittently closer and farther from the line
of sight with an excursion of several degrees in one day, should produce
noticeable variable $\gamma$-ray emission on the same (short) time-scales. The
amplification due to the Doppler factor for Compton scattering of stellar
photons is $\delta^{3-2\alpha}$, and therefore even higher than that for
synchrotron emission, i.e. $\delta^{2-\alpha}$ (Georganopoulos et~al.
\cite{georganopoulos01}; Kaufman Bernad\'o et~al. \cite{kaufman02}). Indeed,
such daily variations have been well established in EGRET observations (Tavani
et~al. \cite{tavani98}; Wallace et~al. \cite{wallace00}) and never understood,
because the other class of galactic gamma-ray sources, namely the pulsars,
have steady properties. These variations can now be well understood:
\object{LS~I~+61$^{\circ}$303} hosts a microquasar with a very fast precessing
jet. At some epoch the jet points directly toward the Earth and is therefore a
micro-blazar (Kaufman Bernad\'o et~al. \cite{kaufman02}, Romero et~al.
\cite{romero02}).

Our discovery, explaining the nature of the enigmatic source
\object{2CG~135+01}, has two important consequences. On the one hand 
\object{LS~I~+61$^{\circ}$303} becomes the ideal laboratory to test the
recently proposed model for microblazars with INTEGRAL and MERLIN
observations, and by AGILE and GLAST in the future.
On the other hand this discovery implies that other variable galactic-gamma
ray sources could also be precessing microquasars. Therefore, identification
of their radio counterparts and subsequent high resolution interferometric
observations could increase the still small ($<20$) number of these
fascinating miniatures of the AGNs, called microquasars.

\begin{acknowledgements}

We thank S.~A. Trushkin for providing the RATAN-600 data.
We are grateful to Karl Menten, Jurgen Neidh\"ofer and Edward Polehampton
for comments and suggestions.
We acknowledge detailed and useful comments from A.~R. Taylor, the referee of this letter.
MERLIN is operated as a National Facility by the University of Manchester at Jodrell Bank Observatory on behalf of the UK Particle Physics \& Astronomy Research Council.
M.~R., J.~M.~P. and J.~M. acknowledge partial support by DGI of the Ministerio de Ciencia y Tecnolog\'{\i}a (Spain) under grant AYA2001-3092, as well as partial support by the European Regional Development Fund (ERDF/FEDER).
M.~R. acknowledges support by a Marie Curie Fellowship of the European Community programme Improving Human Potential under contract number HPMF-CT-2002-02053.
M.~P. acknowledges financial support by the program `Ram\'on y Cajal' of the
Ministerio de Ciencia y Tecnolog\'{\i}a (Spain).
J.~M. has been aided in this work by an Henri Chr\'etien International
Research Grant administered by the American Astronomical Society, and has been
partially supported by the Plan Andaluz de Investigaci\'on of the Junta de
Andaluc\'{\i}a (ref. FQM322).

\end{acknowledgements}

\end{document}